\documentclass[11pt]{article}
\usepackage[T1]{fontenc}    
\usepackage{geometry}       
\usepackage{amsmath}        
\usepackage{amsfonts}       
\usepackage{graphicx}       
\usepackage{float}          
\usepackage{caption}        
\usepackage{subcaption}     
\usepackage{authblk}        
\usepackage[numbers,sort&compress,elide]{natbib} 
\usepackage{hyperref}       
\usepackage{overpic} 

\graphicspath{{figures/}}   
\hypersetup{colorlinks=true,citecolor=blue, linkcolor=blue,urlcolor=blue} 
\date{\today}
\begin{document}
\title{Neutralization of Levitated Charged Nanodiamond: Towards matter-wave interferometry with massive objects}
\author[1]{Sela Liran}
\author[1]{Or Dobkowski}
\author[1]{Rafael Benjaminov}
\author[1]{Peter Skakunenko}
\author[1]{Michael Averbukh}
\author[1]{Yaniv Bar-Haim}
\author[1]{David Groswasser}
\author[2]{Joshua H. Baraban}
\author[1]{Ron Folman\thanks{Corresponding author: \href{mailto:folman@bgu.ac.il}{folman@bgu.ac.il}}}
\affil[1]{Ben-Gurion University of the Negev, Department of Physics, Be'er Sheva 84105, Israel}
\affil[2]{Ben-Gurion University of the Negev, Department of Chemistry, Be'er Sheva 84105, Israel}
\maketitle
\begin{abstract}
Quantum mechanics (QM) and General relativity (GR), also known as the theory of gravity, are the two pillars of modern physics. A matter-wave interferometer with a massive particle, can test numerous fundamental ideas, including the spatial superposition principle - a foundational concept in QM - in completely new regimes, as well as the interface between QM and GR, e.g., testing the quantization of gravity. Consequently, there exists an intensive effort to realize such an interferometer. While several paths are being pursued, we focus on utilizing nanodiamonds as our particle, and a spin embedded in the ND together with Stern-Gerlach forces, to achieve a closed loop in space-time. There is a growing community of groups pursuing this path \cite{whitepaper_cern_xxx}. We are posting this technical note (as part of a series of seven such notes), to highlight our plans and solutions concerning various challenges in this ambitious endeavor, hoping this will support this growing community.
 In this work we demonstrate the neutralization of levitated nanodiamonds using ultraviolet photoemission, and characterize the dependence of this process on both the illumination wavelength and particle size. Furthermore, we demonstrate discrete, single-electron charge manipulation of levitated nanodiamond in a needle Paul trap at a pressure of 0.5\,Torr. Finally, we demonstrate fast neutralization of levitated nanodiamonds, achieving a neutralization rate much faster than the state of the art. As neutralization is crucial to avoid spatial decoherence, this constitutes a significant step towards the realization of a nanodiamond spatial interferometer. We would be happy to make available more details upon request.
\end{abstract}
\section{Introduction}
\subsection{Nanodiamonds Interferometer as a Probe of Quantum Gravity}
Quantum mechanics (QM) is a pillar of modern physics. It is thus imperative to test it in ever-growing regions of the relevant parameter space. A second pillar is general relativity (GR), and as a unification of the two seems to be eluding continuous theoretical efforts, it is just as imperative to experimentally test the interface of these two pillars by conducting experiments in which foundational concepts of the two theories must work in concert.

The most advanced demonstrations of massive spatial superpositions have been achieved by Markus Arndt's group, reaching systems composed of approximately 2,000 atoms\,\cite{Fein2019MoleculeSuperpositions}. This will surely grow by one or two orders of magnitude in the near future. An important question is whether one can find a new technique that would push the state of the art much further in the mass and spatial extent of the superposition. Several paths are being pursued \cite{RomeroIsart2017CoherentInflation, Pino2018OnChipSuperconductingMicrosphere, Weiss2021LargeDelocalizationOptimalControl, Neumeier2024FastQuantumInterference, Kialka2022RoadmapHighMassMWI}, and we choose to utilize Stern-Gerlach forces.

The Stern-Gerlach interferometer (SGI) has, in the last decade, proven to be an agile tool for atom interferometry \cite{Amit2019T3SGInterferometer,Dobkowski2025QuantumEquivalence, Keil2021SternGerlachAtomChip}. Consequently, we, as well as others, aim to utilize it for interferometry with massive particles, specifically, nanodiamonds (NDs) with a single spin embedded in their center \cite{Wan2016FreeRamsey,Scala2013SpinInducedMWI,Margalit2021CompleteSGI}.

Levitating, trapping, and cooling of massive particles, most probably a prerequisite for interferometry with such particles, has made significant progress in recent years. Specifically, the field of levitodynamics is a fast-growing field \cite{gonzalez-ballestero_levitodynamics_2021}. Commonly used particles are silica spheres. As the state of the art spans a wide spectrum of techniques, achievements and applications, instead of referencing numerous works, we take, for the benefit of the reader, the unconventional step of simply mapping some of the principal investigators; these include Markus Aspelmeyer, Lukas Novotny, Peter Barker, Kiyotaka Aikawa, Romain Quidant, Francesco Marin, Hendrik Ulbricht and David Moore. Relevant to this work, a rather new sub-field which is now being developed deals with ND particles, where the significant difference is that a spin with long coherence times may be embedded in the ND. Such a spin, originating from a nitrogen-vacancy (NV) center, could enable the coherent splitting and recombination of the ND by utilizing Stern-Gerlach forces \cite{Margalit2021CompleteSGI, Wan2016FreeRamsey, Scala2013SpinInducedMWI}. This endeavor includes principal investigators such as Tongcang Li, Gavin Morley, Gabriel Hetet, Tracy Northup, Brian D’Urso, Andrew Geraci, Jason Twamley and Gurudev Dutt.

We aim to start with an ND of $10^7$ atoms and extremely short interferometer durations. Closing a loop in space-time in a very short time is enabled by the strong magnetic gradients induced by the current-carrying wires of the atom chip \cite{Keil2016FifteenYearsAtomChip}. Such an interferometer will already enable testing the existing understanding concerning environmental decoherence (e.g., from blackbody radiation), and internal decoherence \cite{HenkelFolman2024UniversalLimitPhonons}, never tested on such a large object in a spatial superposition. As we slowly evolve to higher masses and longer durations (larger splitting), the ND SGI will enable us and others to probe not only the superposition principle in completely new regimes, but in addition, it will enable the community to test specific aspects of exotic ideas such as the Continuous spontaneous localization hypothesis \cite{Adler2021CSLLayering,Gasbarri2021TestingFoundationsSpace}. As the masses are increased, the ND SGI will be able to test hypotheses related to gravity, such as modifications to gravity at short ranges (also known as the fifth force), as one of the SGI paths may be brought in a controlled manner extremely close to a massive object \cite{Geraci2010ShortRangeForceMSpheres,GeraciGoldman2015ShortRangeNanosphereMWI,Bobowski2024ShortRangeAnisotropic,Panda2024LatticeGravAttraction}. Once SGI technology allows for even larger masses ($10^{11}$ atoms), we could test the Diósi–Penrose collapse hypothesis \cite{Penrose2014GravitizationQM,FuentesPenrose2018QuantumStateReductionBEC,Howl2019BECUnification,Tomaz2024CollapseTimeMolecules,Bassi2013CollapseModelsRMP} and gravity self-interaction \cite{HatifiDurt2023HumptyDumpty,Grossardt2021DephasingSemiclassical,AguiarMatsas2024SchrodingerNewtonSG} (e.g., the Schrödinger-Newton equation). Here starts the regime of active masses, whereby not only the gravitation of Earth needs to be taken into account. Furthermore, it is claimed that placing two such SGIs in parallel will allow probing the quantum nature of gravity \cite{Bose2017SpinEntanglementQG,MarlettoVedral2017GravInducedEntanglement}. This will be enabled by ND SGI, as with $10^{11}$ atoms, the gravitational interaction could be the strongest \cite{VanDeKamp2020CasimirScreening,Schut2023RelaxationQGIM,Schut2024MicronSizeQGEM}.

Let us emphasize that, although high accelerations may be obtained with multiple spins, we consider only a ND with a single spin, as numerous spins will result in multiple trajectories and will smear the interferometer signal. We also note that working with an ND with less than $10^7$ atoms is probably not feasible because of two reasons. The first is that NVs that are closer to the surface than 20\,nm lose coherence, and the second is that at sizes smaller than 50\,nm, the relative fabrication errors become large, and a high-precision ND source becomes beyond reach.

Here we demonstrate neutralization of NDs using ultraviolet (UV) photoemission, and characterize the dependence of this process on both the illumination wavelength and particle size. Furthermore, we demonstrate discrete, single-electron charge manipulation of levitated nanodiamond in a needle Paul trap at a pressure of 0.5\,Torr. Finally, we demonstrate fast neutralization of levitated nanodiamonds, using a focused UV laser beam, achieving a neutralization rate faster than the state of the art by at least an order of magnitude. This technical note is part of a series of seven technical notes put on the archive towards the end of August 2025, including a wide range of required building blocks for a ND SGI \cite{Muretova_ND_theory,Givon_ND_fabrication,Benjaminov_ND_loading,Feldman_Paul_trap_ND,skakunenko_strong_nodate,Levi_Quantum_control_NV}.

\subsection{Neutralization of Levitated Charged Nanodiamonds}
\paragraph{}
For a spatial matter-wave interferometer with massive particles to work, one has to suppress numerous spatial decoherence channels. Blackbody radiation emitted from the environment or from the particle itself is one such channel. Suppressing it requires cooling the internal temperature of the particle and the environment (e.g., the vacuum chamber)~\cite{HenkelFolman2024UniversalLimitPhonons}. In this work, our main interest is the strong coupling of a charged particle to its environment. Such a charge on the particle can couple strongly to electromagnetic fields in the environment, and give rise to significant decoherence rates, which degrade the interference signal\,\cite{sonnentag_measurement_2007,scheel_path_2012}.
To avoid this, one must neutralize the particle. Another undesired effect of excess charge on the surface of the particle is the suppression of the spin coherence time of the NV center embedded in the ND. It has been demonstrated that excess charge on the surface of NDs shortens the spin coherence time, especially for very small NDs, where the NV center is very close to the surface\,\cite{kim_decoherence_2015}. In conclusion, neutralization of trapped NDs is a vital step towards the realization of a matter-wave interferometer with NDs.

Neutralization and precise control of electric charge on particles in optical or magneto-gravitational traps has been demonstrated using several methods, including corona discharge~\cite{frimmer_controlling_2017, kamba_optical_2022}, exposure to alpha radiation~\cite{slezak_cooling_2018, marmolejo_visualizing_2021}, and ultraviolet (UV) illumination~\cite{moore_search_2014, monteiro_force_2020, slezak_cooling_2018, kamba_optical_2022}. These methods typically enable the observation of discrete single-electron charge change.

Corona discharge neutralization was demonstrated by Frimmer et al.~\cite{frimmer_controlling_2017} at pressures of 0.1--1\,mbar, enabling single-electron charge control of optically levitated silica nanoparticles (136\,nm), with a characteristic timescale of approximately 10 seconds per electron. Neutralization by alpha radiation emitted from \textsuperscript{241}Am was demonstrated by Slezak et al.~\cite{slezak_cooling_2018} on magneto-gravitationally trapped silica microspheres (5\,\textmu m) at atmospheric pressure. The neutralization rate showed a characteristic timescale of approximately one second per electron. In that same work, the last neutralization step was completed in high vacuum using a UV lamp, which demonstrated a characteristic timescale of about 100 seconds per electron. This method was also demonstrated to roughly neutralize a drop of liquid in which a nanodiamond was suspended in a trapped droplet, which later evaporated, leaving the nanodiamond trapped~\cite{hsu_cooling_2016}. Another use of charge control by alpha radiation~\cite{marmolejo_visualizing_2021} showed discrete charge manipulation, and visualized it as stepwise shifts in the vertical position of a levitated silicone oil droplet (30\,\textmu m), making the discrete charge change visible to the naked eye.

Neutralization of silica spheres using UV illumination was demonstrated by Moore et al.~\cite{moore_search_2014} under high vacuum conditions ($\sim$10$^{-7}$\,mbar), enabling single-electron charge control of silica microspheres (5\,\textmu m), with a characteristic timescale of approximately 1000 seconds per electron. Monteiro et al.~\cite{monteiro_force_2020} demonstrated UV-induced neutralization of ($\sim$10\,\textmu m) silica microspheres in ultra-high vacuum, achieving a faster charge variation rate of approximately 0.1 seconds per electron. A hybrid neutralization approach combining corona discharge at moderate pressure (3.5\,mbar) and UV illumination at high vacuum (<10$^{-6}$\,mbar) was demonstrated by Kamba et al.~\cite{kamba_optical_2022}, achieving single-electron charge control of silica nanoparticles (166\,nm), with a characteristic timescale of approximately 10 seconds per electron.

Current works on single-electron charge control show a wide range of neutralization timescales ranging between a tenth of a second \cite{monteiro_force_2020} and thousands of seconds per electron \cite{moore_search_2014}. A typical nanoparticle trapped in a Paul trap has excess charge on the order of tens to 100 elementary charges. It is favorable for the neutralization process to be short, much less than a second, so that the particle does not heat internally or externally. Ultimately, we would like to reach a duration of milliseconds per electron. In summary, while neutralization at a rate of 0.1 seconds per electron is assumed to be fast enough for the first realizations of an ND SGI, we would like to reach a neutralization time of milliseconds per electron. In addition, the above result was not achieved for an ND and this warrants further development.
Furthermore, while corona discharge and alpha radiation methods require residual gas interactions, UV photoemission enables neutralization under ultra-high vacuum conditions, a critical advantage for quantum interferometry applications. Future developments, particularly the implementation of focused UV laser radiation, promise to deliver the high photon flux densities necessary for achieving the desired rapid neutralization rate. Indeed, in this work we show that focusing such a laser enables us to go beyond the state of the art, addressing all the above problems.

\subsection{Photoemission from Diamonds}
UV photoemission has been demonstrated in levitated nanoscale silica spheres, but, to the best of our knowledge, not yet in levitated NDs. Photoemission from bulk diamonds and NDs is well established and has been applied in various fields, such as electron microscopy \cite{tafel_femtosecond_2019}. However, electric charge neutralization has not yet been shown for levitated NDs, although neutralization of NV centers in NDs has been reported \cite{dhomkar_demand_2018, pederson_rapid_2024}. Consequently, the photoemission properties of levitated NDs remain largely unexplored. In particular, the threshold wavelength for photoemission—a key parameter for UV-based neutralization—has not yet been measured. This value, which is associated with the work function of diamonds, may vary depending on surface conditions, for instance, Tafel et. al \cite{tafel_femtosecond_2019} demonstrated electron emission from ND-coated tungsten tips using femtosecond lasers, reporting a one-photon photoelectric effect at 4.8\,eV (256\,nm) and multiphoton processes in the range of 2.8--4.8\,eV. Y. Zhang et. al~\cite{Zhang_Abs_2023}, characterized the absorption of diamonds with NV centers, showing a threshold in absorption around 5.4\,eV~(230\,nm). Other studies have reported work function values of 4.2\,eV~(295\,nm) \cite{mackie_work_1996} and 5.5\,eV~(225\,nm) \cite{velardi_uv_2017}, with significantly reduced quantum efficiency above 5.8\,eV~(210\,nm). These findings highlight the importance of characterizing the photoemission from levitated NDs, especially in the case of neutralization with laser radiation, in which the wavelength of the laser must be below the threshold wavelength, to achieve efficient neutralization. The specific conditions of the ND surface effect not only the photoemission, but also effect the NV center's quantum properties. For example, it was demonstrated that surface termination can give significant improvement in the optical stability and quantum coherence of NV centers. \cite{malkinsonEnhancedQuantumProperties2024}

\subsection{The Source of Emitted Electrons}
One might assume that during the neutralization of the ND only the excess electrons are removed from the ND surface, but this is not necessarily the case, as it is also possible to eject electrons from the ND bulk. (The UV can also interact with the NV defect centers embedded in the ND, which can lead to the emission of electrons from these centers~\cite{Razinkovas_2021}.) Previous works have shown that removal of excess charge can be achieved by an energy smaller than the work function~\cite{ussenovLaserstimulatedPhotodetachmentElectrons2024}, and more specifically, that excess triboelectric charge release has a lower threshold energy than the work function~\cite{linElectronTransferNanoscale2019}. These two distinct mechanisms (surface photodetachment and bulk photoionization) can occur at different threshold wavelengths with potentially significantly different neutralization efficiencies and outcomes, in terms of the resulting state and internal temperature of the ND. The measurement and analysis of the two mechanisms are beyond the scope of this work. While the methods presented here may serve as a potent tool to test the two mechanisms, we leave this question open for future investigations.
\section{Experimental Setup}
\begin{figure}[H]
  \centering
  \includegraphics[width=0.65\textwidth]{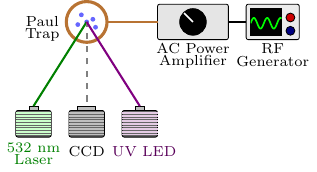}
   \caption{Schematic diagram of the experimental setup. NDs are trapped in a copper ring Paul trap with an inner diameter of 6\,mm. The trap is driven by a homemade high-voltage AC amplifier (designed and built by our local electronics engineer), which amplifies the signal of an RF generator (UNI-T signal generator). We used a range of 2.5--6 kV P-P, and frequency in the range of 100--200\,Hz. A 532\,nm laser is used for particle detection, with scattered light collected by a CCD camera (IDS Ueye). UV LEDs at various wavelengths in the range of 255--310\,nm are focused onto the trap region using fused silica lenses to enhance photoemission.}
  \label{fig:experimental_setup}
\end{figure}

\paragraph{}
The experimental setup, schematically shown in Fig.\,\ref{fig:experimental_setup}, is composed of four main components: a ring Paul trap for trapping NDs, a high voltage AC power supply to drive the trap, an optical detection system for monitoring the particles, and a UV-based neutralization system. Each component is briefly introduced in Fig.\,\ref{fig:experimental_setup} and described in detail in the following subsections.
\subsection{Nanodiamond Particles}
The ND particles used in this study were obtained from \href{https://pureon.com/products/micron-diamond-powders/}{Pureon} and synthesized using high-pressure high-temperature (HPHT) methods. The supplied powder included particles with diameters ranging from 75\,nm to 10\,$\mu$m. Prior to being introduced into the experimental setup, the NDs were dried in an oven at 100\,$^\circ$C for three hours to remove residual moisture.

The size range of the NDs for these measurements was chosen to match the predicted size range in the ND interferometer, in which we plan to use NDs containing $10^7$ to $10^{11}$ atoms. To estimate the size of such particles, we assume a spherical geometry, and given the density of diamond ($3.52 \times 10^3$ kg/m$^3$) and the atomic mass of carbon (12.01 g/mol), we calculate the volume per atom to be $5.67 \times 10^{-30}$ m$^3$. Consequently, spherical NDs within our targeted atom-count range correspond to radii between roughly 25\,nm ($10^7$\,atoms) and 500\,nm ($10^{11}$\,atoms).

\subsection{UV Source}
The UV source used in this experiment consists of LEDs from multiple manufacturers (\href{https://lianxinrui.en.alibaba.com/}{Lianxinrui} and \href{https://www.uvwavetek.com/}{UVWavetek}). These LEDs have designed wavelengths of 255, 265, 280, and 310\,nm, with specified spectral widths of approximately 5\,nm and optical power ranging from 1 to 15\,mW. The actual wavelengths and spectral characteristics were characterized using a Flame spectrometer from \href{https://www.oceanoptics.com/wp-content/uploads/2024/12/Flame-User-Manual.pdf}{Ocean Optics}, with results reported in Figs.\,\ref{fig:decay_fits}--\ref{fig:freq_vs_UVtime}. Fused silica lenses are used to focus the UV light into a concentrated beam directed at the copper ring of the Paul trap. However, due to conservation of étendue~\cite{etendue_conversion}, the intensity at the focal point cannot exceed the source intensity of the LED itself, which fundamentally limits the maximum achievable photon flux density and puts a boundary on the neutralization rate. This, in contrast to a laser beam, which can be tightly focused to increase its intensity, emphasizes the need for a UV laser based neutralization as a next step. Focusing the beam on the plane of the trap (rather than collimating the beam) has an advantage, as it reduces the intensity impinging on distant surfaces, thereby avoiding unintended photoemission effects from such surfaces. 

\subsection{Particle Detection}
\begin{figure}[H]
  \centering
  \begin{subfigure}[b]{0.24\textwidth}
    \includegraphics[width=\textwidth]{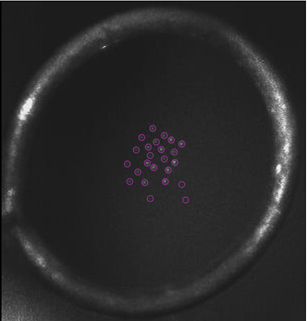}
    \caption{t = 0 s, N = 19}
    \label{fig:particle_detection_a}
  \end{subfigure}
  \hfill
  \begin{subfigure}[b]{0.24\textwidth}
    \includegraphics[width=\textwidth]{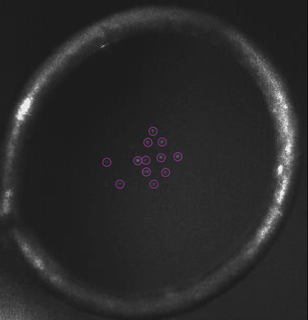}
    \caption{t = 18 s, N = 12}
    \label{fig:particle_detection_b}
  \end{subfigure}
  \hfill
  \begin{subfigure}[b]{0.24\textwidth}
    \includegraphics[width=\textwidth]{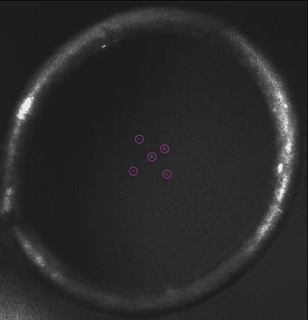}
    \caption{t = 60 s, N = 5}
    \label{fig:particle_detection_c}
  \end{subfigure}
  \hfill
  \begin{subfigure}[b]{0.24\textwidth}
    \includegraphics[width=\textwidth]{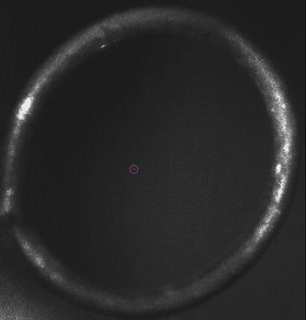}
    \caption{t = 100 s, N = 1}
    \label{fig:particle_detection_d}
  \end{subfigure}
  \caption{CCD camera frames of the particles trapped in the ring Paul trap. The white bright spots correspond to detected particles, automatically marked by the TrackMate algorithm by a purple circles.
  }
  \label{fig:particle_detection_combined}
\end{figure}

To count the trapped particles, we illuminate them with a green laser and detect the light scattered from the diamonds using a CCD camera and record it as a video sequence. The recordings are then analyzed using the TrackMate algorithm \cite{tinevez_trackmate_2017} in the FIJI software platform. This program automatically identifies the bright spots generated by the scattered laser light from individual ND particles.  An example of the particle detection process is shown in Fig.\,\ref{fig:particle_detection_combined}.

\subsection{Paul Trap}
To trap the ND particles, we constructed a ring Paul trap, which is a type of quadrupole trap that uses oscillating electric fields to confine charged particles. The design of the trap is based on the principles of dynamic stabilization, where the time-varying electric fields create a potential well that can confine particles in three dimensions. The core of the Paul trap is a homemade copper ring electrode, with an inner diameter of 6\,mm. A high-voltage potential is applied to the ring, supplied by a homemade high-voltage AC amplifier (designed and built by our local electronics engineer), which amplifies the signal from an RF generator (UNI-T signal generator). We used a range of 2.5--6\,kV P-P, with a frequency in the range of 100--200\,Hz. It is optimized for the capacitive load of the ring electrode and incorporates robust insulation and safety features to safely handle high voltages.
The quadrupole potential at the trap center, with negligible DC voltage and an RF voltage $V\cos(\Omega t)$, can be expressed as shown in~\cite{ghosh_ion_traps} 
\begin{equation}
    q = \frac{2 Q V \eta}{m \Omega^2 r_0^2},
    \label{eq:Mathieu_stability_parameter}
\end{equation}
where \(q\) is the stability parameter, \(Q\) is the charge of the particle, \(m\) is the mass of the particle, \(V\) is the amplitude of the RF voltage, \(\Omega\) is the angular frequency of the RF field, \(\eta\) is a geometric factor that depends on the trap geometry, and \(r_0\) is the characteristic distance from the trap center to the electrodes.
The stability range in an atmospheric environment, as reported in \cite{ghosh_ion_traps}, is between 0.1 to 0.9. This condition ensures that the particles remain confined within the trap and do not escape due to instabilities in the potential well. Those trap parameters and particle mass will help us to estimate the minimum and maximum number of elementary charges on the particles that are trapped. 

\subsection{Particle Charging and Loading to the Paul Trap}
Particles are electrically charged by placing them on an ITO-coated glass slide connected to a DC voltage source, and operating the ring Paul trap at a small distance from the slide. Turning on the Paul trap launches particles into the trap, where they levitate in a stable manner. This method of direct electrical charging, launching and loading of NDs into the Paul trap, in ambient conditions, shows high efficiency and high success rate. We detail this method in a paper soon to be published \cite{Benjaminov_ND_loading}. We also found that the slide’s polarity defines the sign of the acquired charge on the NDs, which allows us to test the effect of the UV light for both positively and negatively charged particles.

\begin{figure}[H]
  \centering
  \includegraphics[width=\textwidth]{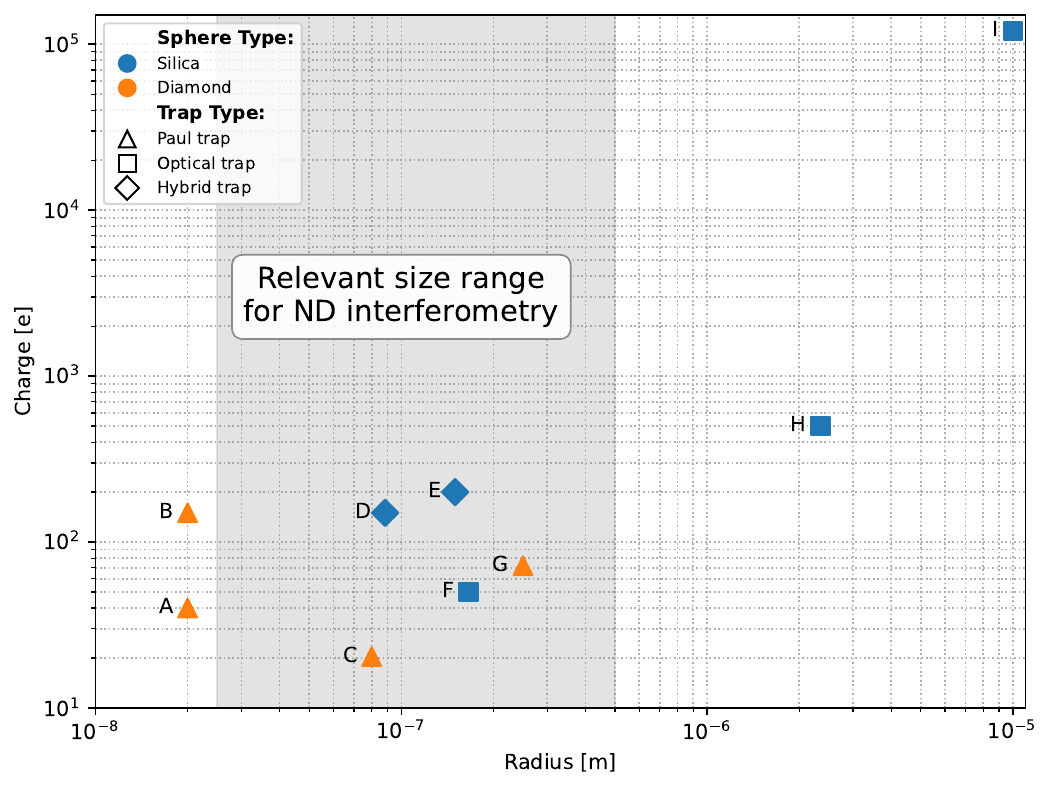}
  \caption{Charging of particles (in units of elementary charge) vs radius, for diamond and silica particles, confined in various traps. Silica spheres are shown in blue, and ND measurements are shown in orange. The data points were extracted from various works, including two from our group: A and B from~\cite{Conangla2020_PhDThesis}, C from~\cite{skakunenko_strong_nodate}, D from~\cite{Bonvin2023_HybridPaulOpticalTrap}, E from~\cite{Bykov2019_DirectLoadingPaul}, F from~\cite{kamba_optical_2022}, G is a measurement from this work, H from~\cite{Blakemore2019_MassDensityMSphere}, and I from~\cite{monteiro_force_2020}. The gray shaded region indicates the particle size range relevant for our first planned matter-wave interferometry experiments (about $10^7$ atoms).}
  \label{fig:charge_vs_radius}
\end{figure}

Estimating the number of elementary charges on the levitated particle is crucial for the neutralization process. In Fig.\,\ref{fig:charge_vs_radius} we show typical values of the number of elementary charges as a function of particle radius, where the values are compiled from data reported for both silica and diamond particles confined in Paul traps, optical tweezers, and hybrid Paul–optical systems~\cite{Conangla2020_PhDThesis, Bonvin2023_HybridPaulOpticalTrap, Bykov2019_DirectLoadingPaul, kamba_optical_2022, Blakemore2019_MassDensityMSphere, monteiro_force_2020, skakunenko_strong_nodate}.

From the charge-to-size ratio of particles that are shown in Fig.\,\ref{fig:charge_vs_radius}, we can estimate the minimum and maximum number of elementary charges for the particles used in this work, which range between 10 electrons for the smallest particles (75\,nm) and 1000 electrons for the largest particles (10\,$\mu$m). These estimates provide a baseline for understanding the initial charge conditions of the particles before neutralization experiments.

\section{Results}
To characterize the photoemission of electrons from trapped NDs and microdiamonds, we developed a method based on lifetime measurements of particles within the trap. As the stability of a particle in the trap strongly depends on its charge-to-mass ratio, charging or discharging the particle due to electron photoemission will destabilize particles, causing them to escape from the trap, exhibiting a lifetime that is dependent on the photoemission rate. This method proved efficient and straightforward, allowing us to measure the effect of wavelength and particle size on the photoemission process.

\begin{figure}[H]
  \centering
    \begin{overpic}[width=\textwidth]{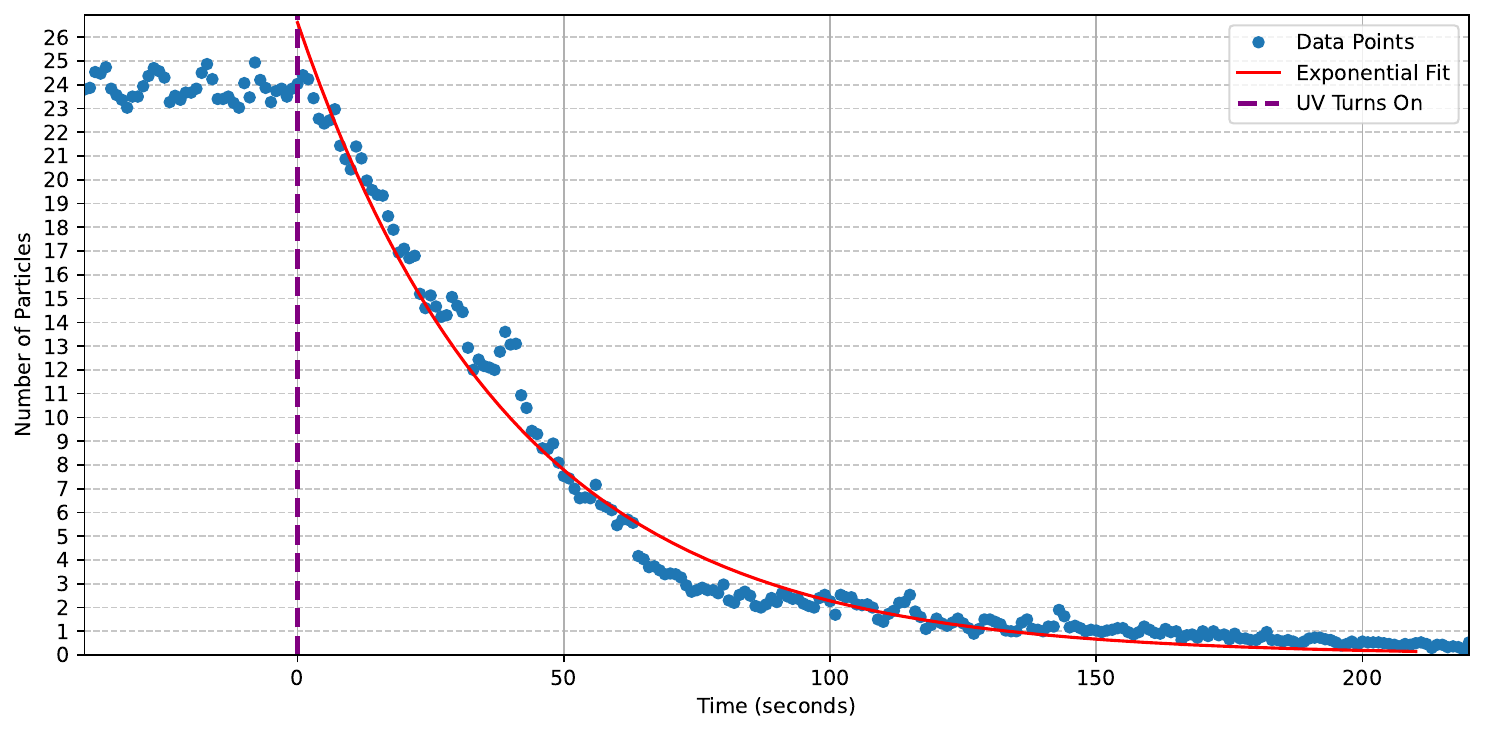}
      \put(48,35){
        \begin{minipage}{0.5\textwidth}
        \includegraphics[width=\linewidth]{{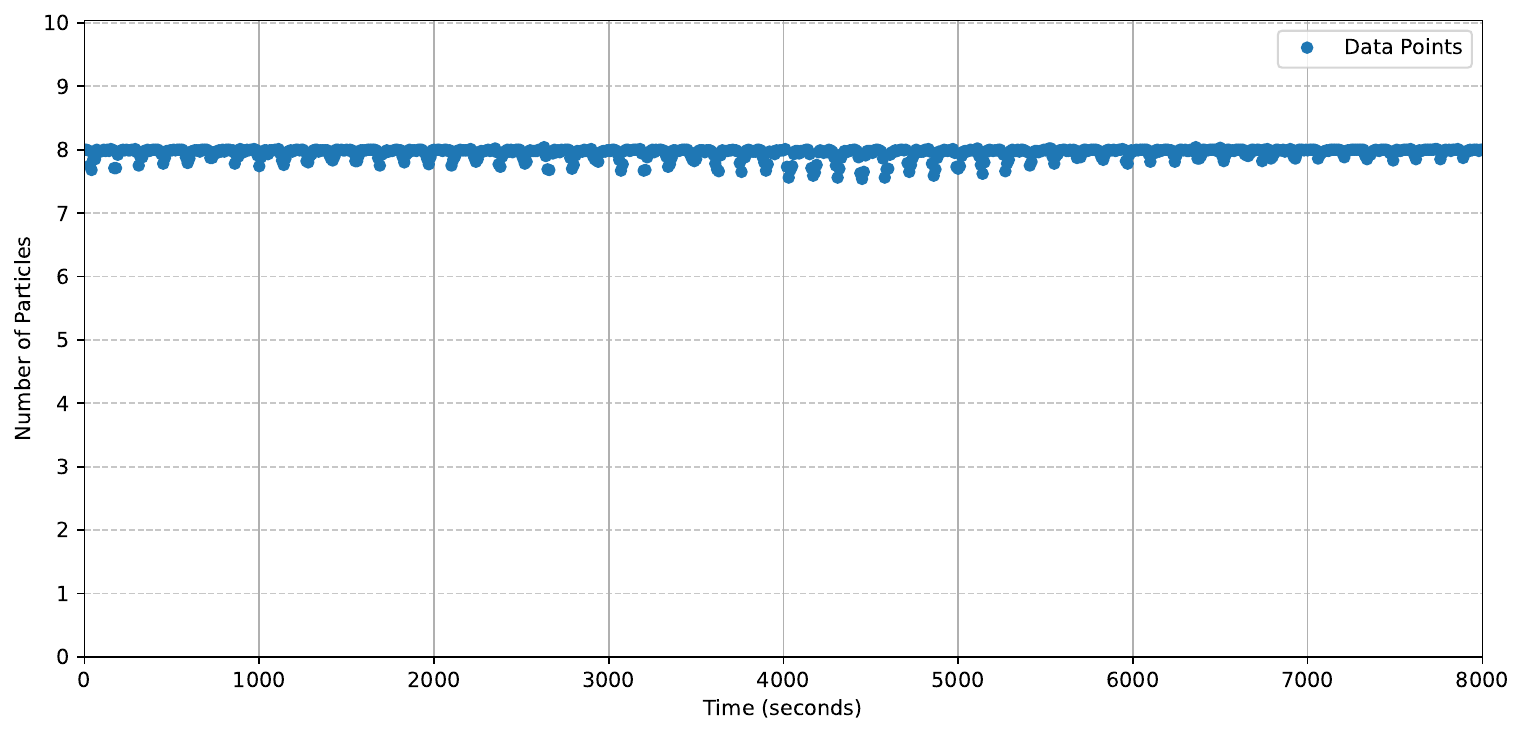}}
        \end{minipage}
      }
  \end{overpic}
  \caption{Number of particles vs time, demonstrating the decay in trapped particle count under UV illumination with a $263\pm 5$\,nm LED. The data are fitted with an exponential decay model (Eq.~\ref{eq:exponential_decay}), yielding a characteristic decay time of $\tau=40.7\pm0.4$\,s. The inset presents a control measurement without UV illumination, where particles remained trapped for over 8,000 seconds.}
  \label{fig:decay_fits}
\end{figure}

Initially, we verified that trapped particles remain stable without illumination, exhibiting lifetimes on the order of hours or longer (see the inset in Fig.\,\ref{fig:decay_fits}). Subsequently, we exposed negatively charged particles to UV illumination, observing significantly shorter lifetime, on the order of a few seconds to hundrnds of seconds (see Figs.\,\ref{fig:decay_fits},\ref{fig:uv_wavelength_subfigures}). We also conducted experiments on positively charged particles under UV illumination and observed longer lifetimes, which further confirmed that particle loss is primarily due to electron photoemission. For each experimental run, we tracked the number of trapped particles over time. The lifetime was determined by fitting the decay of particle count to the an exponential function:
\begin{equation}
N(t) = N_0 \cdot e^{-t / \tau}
\label{eq:exponential_decay}
\end{equation}
where $N(t)$ represents the number of trapped particles at time $t$, $N_0$ is the initial particle number, and $\tau$ is the characteristic lifetime. An example of this decay process is shown in Fig.\,\ref{fig:decay_fits}.

As shown in inset of Fig.\,\ref{fig:decay_fits}, to reliably distinguish UV-induced neutralization from other loss mechanisms we performed control measurements. Specifically, we measured the trap lifetime of ND particles without any UV illumination, where under these conditions, the particles remained trapped for extended periods, typically on the order of several hours, confirming that particle loss due to background effects is negligible on the timescale relevant to our experiments.

In the next sections we use this method to study UV-induced photoemission from levitated NDs. In Sec.\,\ref{sec:photoemission_vs_wavelength} we investigate how the illumination wavelength effects the photoemission process. In Sec.\,\ref{sec:particle_size_effects} we study how the particle size affects the photoemission process. In Sec.\,\ref{sec:single_charge_manipulation} we demonstrate another method of measurement, where by monitoring the frequency of oscillations in the Paul trap, we demonstrate discrete single-electron charge manipulation of a levitated ND in a needle Paul trap.

\subsection{Photoemission vs. Wavelength}\label{sec:photoemission_vs_wavelength}
\begin{figure}[H]
  \centering
  \begin{subfigure}[b]{0.48\textwidth}
    \includegraphics[width=\textwidth]{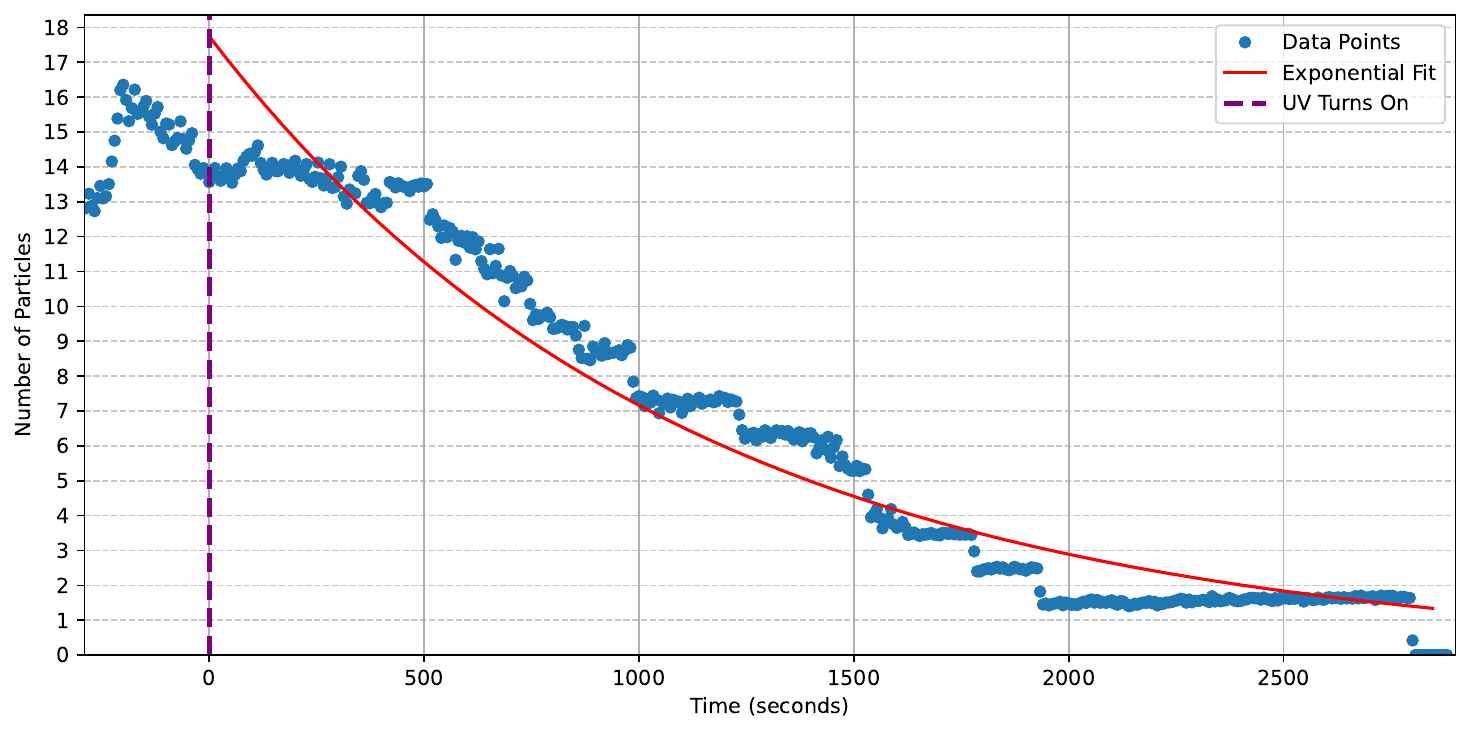}
    \caption{315\,nm}
    \label{fig:310_1mu}
  \end{subfigure}
  \hfill
  \begin{subfigure}[b]{0.48\textwidth}
    \includegraphics[width=\textwidth]{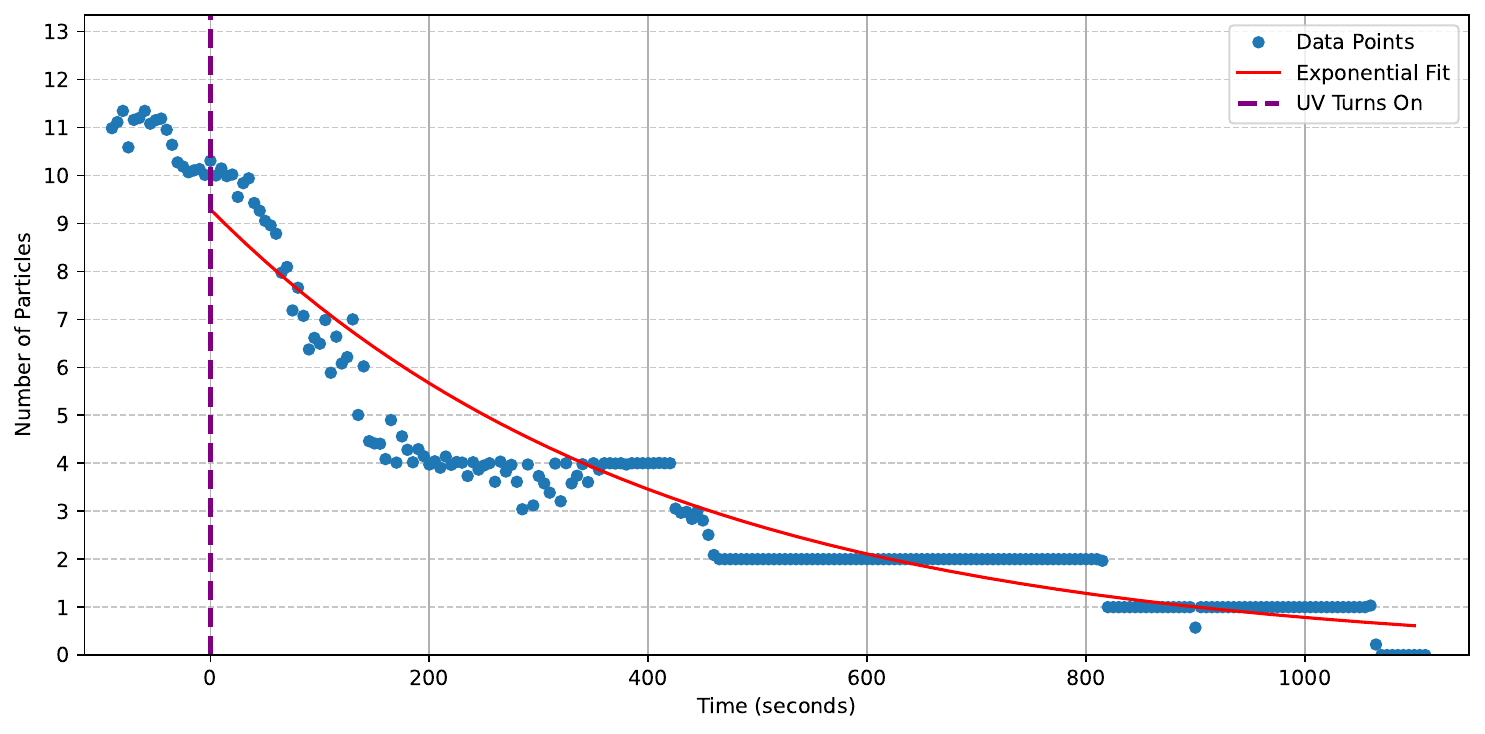}
    \caption{279\,nm}
    \label{fig:280_1mu}
  \end{subfigure}
  \\[1ex]
  \begin{subfigure}[b]{0.48\textwidth}
    \includegraphics[width=\textwidth]{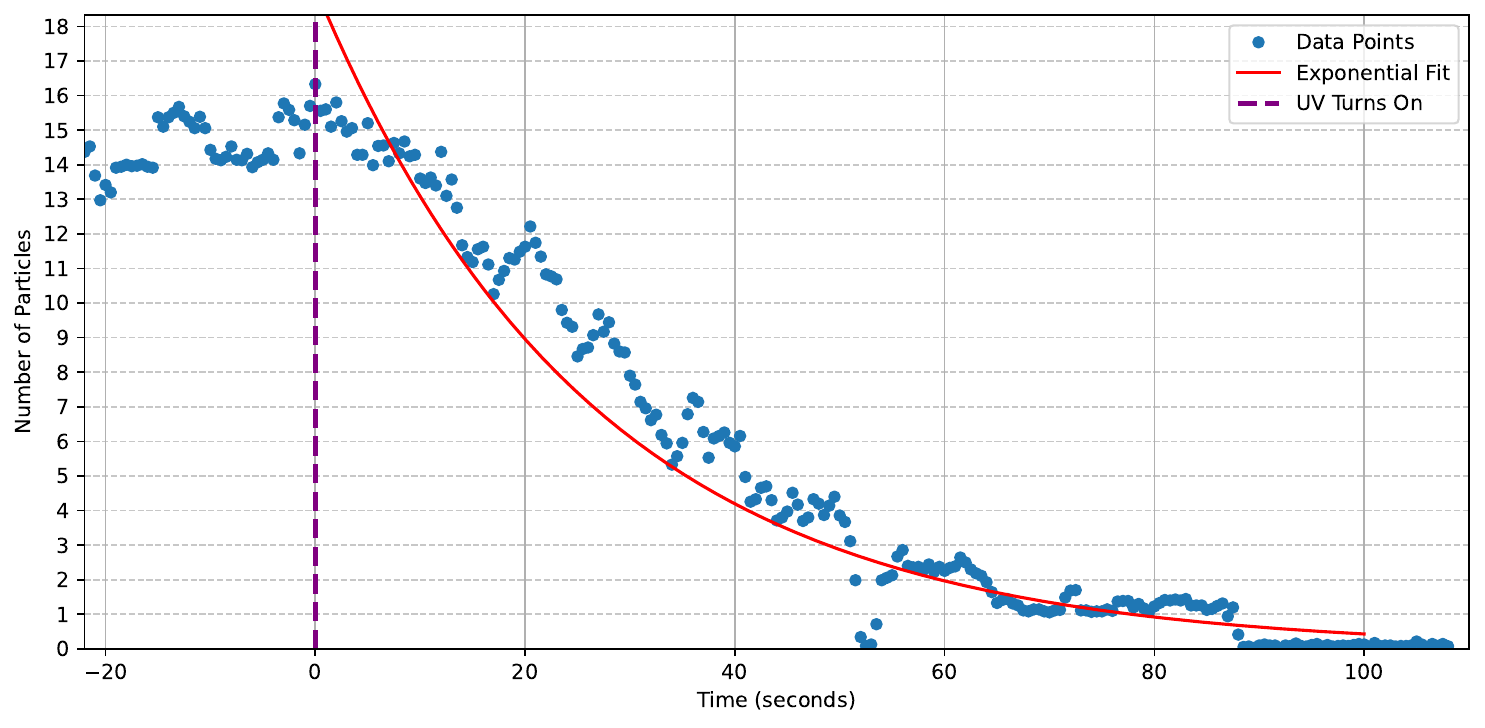}
    \caption{271\,nm}
    \label{fig:265_1mu}
  \end{subfigure}
  \hfill
  \begin{subfigure}[b]{0.48\textwidth}
    \includegraphics[width=\textwidth]{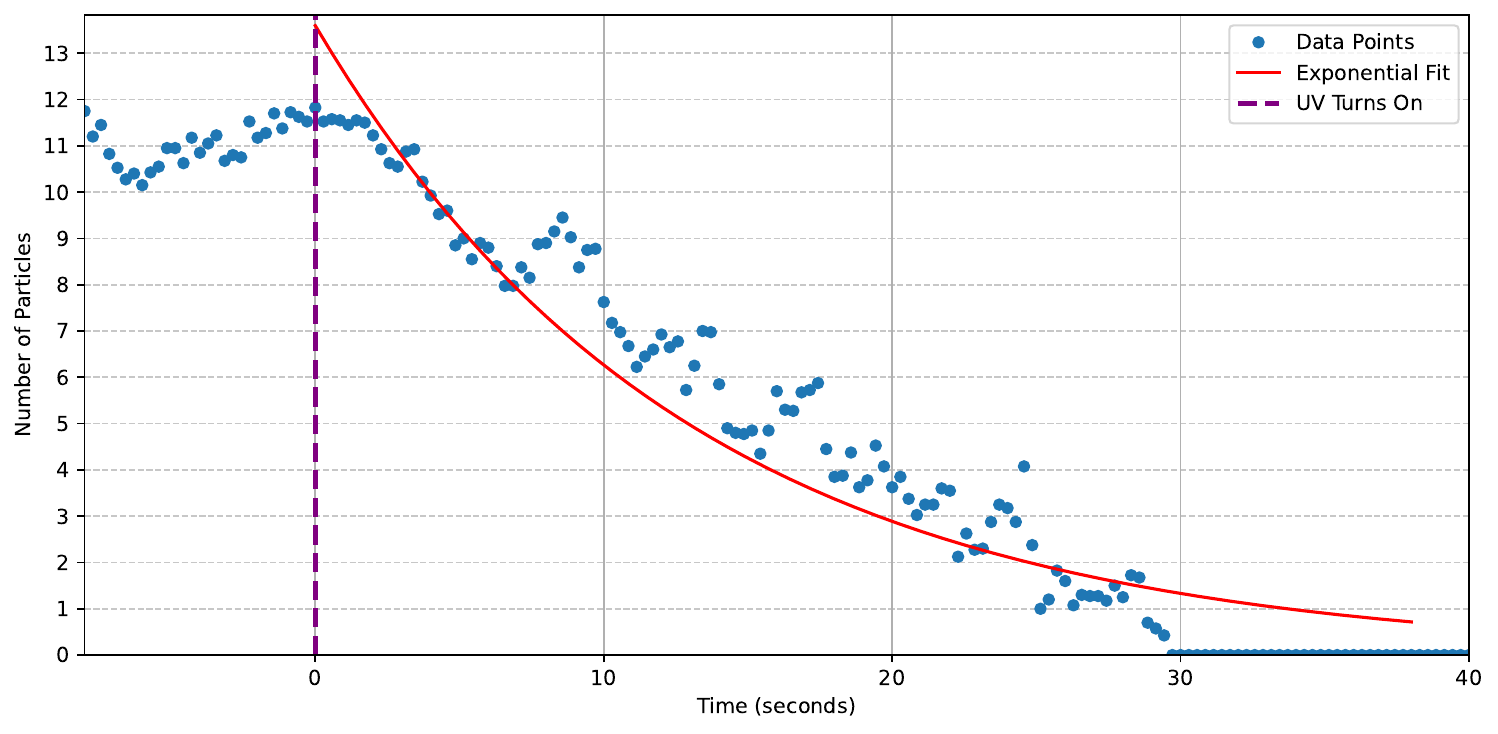}
    \caption{264\,nm}
    \label{fig:250_1mu}
  \end{subfigure}
  \caption{Lifetime measurements of levitated NDs in a Paul trap under UV illumination of various wavelengths. We plot the number of trapped particles as a function of time under wavelengths (a) $315\pm 7$\,nm, (b) $279\pm 6$\,nm, (c) $271\pm 6$\,nm, and (d) $264 \pm 5$\,nm. The onset of UV illumination is indicated in each plot by a vertical dashed purple line. All data sets were taken with 1\,$\mu$m Pureon NDs, trap driving at 140\,Hz and 4.5\,kV P-P, with particles negatively charged by ITO slide held at a negative voltage of \( -150 \,\mathrm{V} \). The optical lenses and LED power were kept constant for all measurements.}
  \label{fig:uv_wavelength_subfigures}
\end{figure}

\begin{figure}[H]
  \centering
  \includegraphics[width=\textwidth]{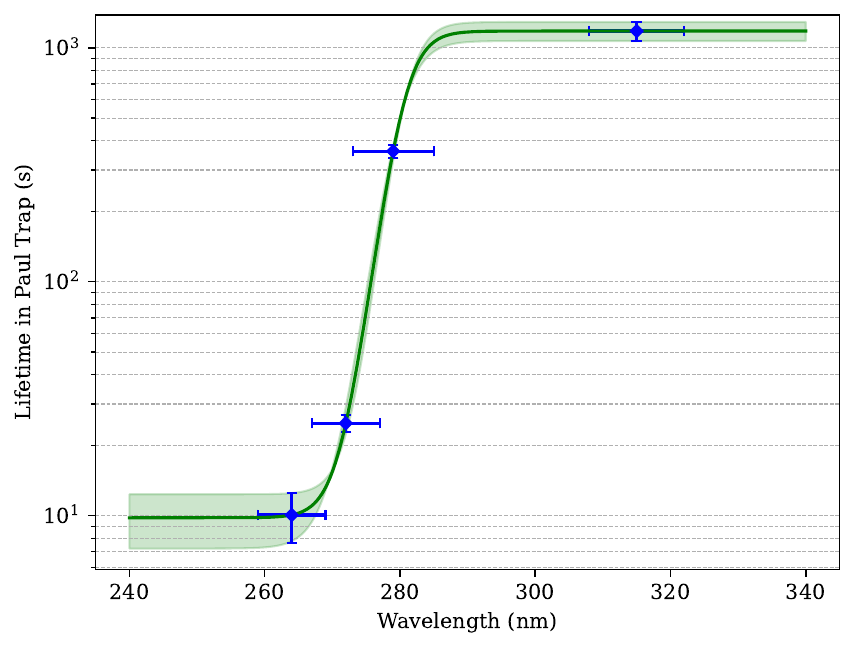}
  \caption{Threshold wavelength for photoemission from levitated ND: Plotting the lifetime of negatively charged NDs as a function of UV illumination wavelength (y-axis in log scale). The points were produced by measurements as in Fig.\,\ref{fig:uv_wavelength_subfigures}, where the vertical error bars indicate the standard error from multiple independent measurements and the horizontal error bars represent the spectral bandwidth of the UV LEDs (5--7\,nm). A distinct threshold behavior is observed where the center of the sigmoid step is at 280\,nm, with a width of 10\,nm (10\% to 90\%), so that the threshold for high efficiency photoemission is at 270\,nm. Below this threshold, photoemission becomes highly efficient, resulting in rapid particle neutralization and dramatically reduced trap lifetimes. The measurements are fitted using a sigmoid function of the form $f(x) = \frac{L}{1 + \exp[-k(\lambda - \lambda_0)]} + b$, where $L$ is the asymptotic maximum, $\lambda_0$ is the threshold, $k$ is the slope, and $b$ is the baseline. Fitting was performed by weighted nonlinear least-squares regression, with data uncertainties used as weights. The shaded region denotes the $1\sigma$ confidence interval from the parameter covariance. This model captures the threshold-like transition in the system's response to illumination wavelength.}
  \label{fig:lifetime_vs_wavelength}
\end{figure}

To determine the photon energy required to eject excess negative charge from NDs, we illuminated trapped particles in the Paul trap with UV light of varying wavelengths. As the photon energy increases (i.e., wavelength decreases), the probability of photoemission rises, eventually leading to loss of charge and destabilization of the trap. In Fig.\,\ref{fig:uv_wavelength_subfigures} we show measurements of trap lifetime as function of wavelength in the range of 264-315\,nm for negatively charged NDs. The results show that the lifetime changes dramatically from a lifetime on the order of a few seconds to a lifetime on the order of more than a thousand seconds, as the wavelength is increased from 264\,nm to 315\,nm. These results are summarized in Fig.\,\ref{fig:lifetime_vs_wavelength}, where we observe a distinct threshold behavior. As the wavelength decreases below this threshold, the trap lifetime drops sharply, signifying efficient photoemission and rapid neutralization of the trapped particles. The threshold corresponds to a photon energy of about 4.6\,eV, consistent with the reported work function values for diamond surfaces (see Fig.\,3. in \cite{tafel_femtosecond_2019})

\subsection{Photoemission vs. particle size}\label{sec:particle_size_effects}
\begin{figure}[H]
    \centering
    \includegraphics[width=\textwidth]{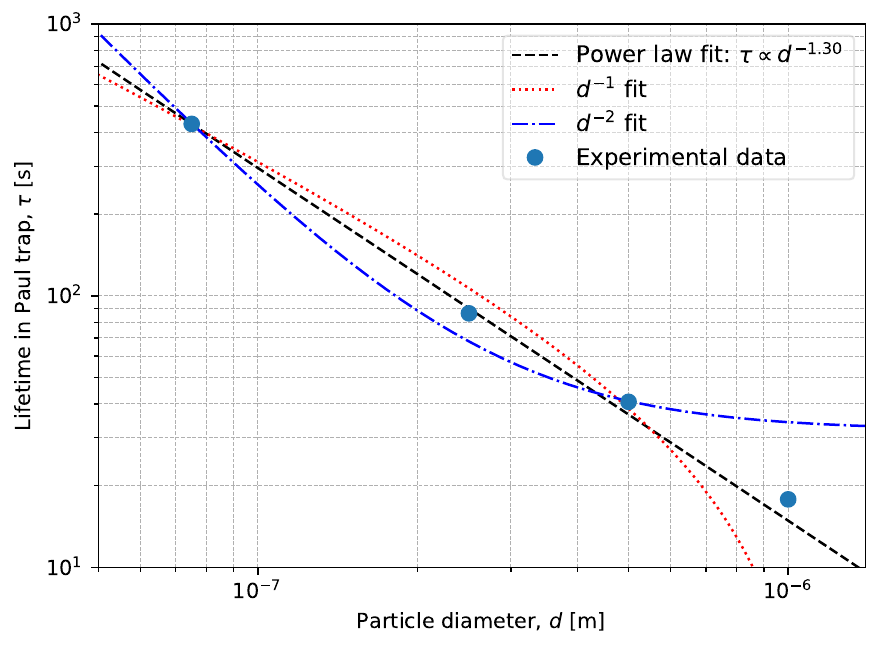}
    \caption{
    Trap lifetime of levitated NDs as a function of particle diameter, displayed on log--log axes. Experimental measurements (circles) show the measured lifetime ($\tau$) for NDs under UV illumination with $264\pm 5$\,nm LED in a ring Paul trap at atmospheric pressure. The black dashed line indicates a power-law fit of the form $\tau \propto d^{-1.3}$. The red dotted and blue dash-dotted lines correspond to fits of $d^-1$ and $d^-2$ dependence, respectively, motivated by surface- and volume-dominated neutralization scenarios. This figure demonstrates the strong dependence of neutralization dynamics on ND size, and is the first step in a much-needed comparison between different scaling models for photoemission-driven discharge.
    }
    \label{fig:lifetime_vs_size}
\end{figure}

To characterize the neutralization dynamics across different particle sizes, we measured the trap lifetime as a function of particle diameter. The results shown in Fig.\,\ref{fig:lifetime_vs_size} show that larger particles are neutralized more rapidly than smaller ones. This size dependence can provide insight into the underlying neutralization mechanisms and has important implications for optimizing fast neutralization protocols.

From these results, we can estimate the requirements for fast neutralization for the ND SGI. The relevant ND size range for matter-wave interferometry applications is 25--500\,nm (see Fig.\,\ref{fig:charge_vs_radius}). Our measurements in Fig.\,\ref{fig:lifetime_vs_size} show that neutralization of a 75\,nm ND required a duration of approximately 500\,s under LED illumination at an intensity of $\sim$1\,mW/cm$^2$. To achieve a similar neutralization process in 10\,ms would require increasing the optical intensity by a factor of $\sim 50,000$. This intensity level is readily achievable with focused UV laser radiation. For a laser with a total power of 10\,mW, it requires focusing the beam to a diameter of $\sim \sqrt{1/5000}\,\mathrm{cm}\approx 150\,\mathrm{\mu m}$, which is feasible with one standard lens in the path of a collimated laser beam. This analysis suggests that rapid neutralization of levitated NDs is achievable with a $10\,\mathrm{mW}$ focused 266\,nm laser, which is within the capabilities of current laser technology.

\subsection{Single Charge Manipulation of Levitated Charged Nanodiamonds using UV Photoemission}\label{sec:single_charge_manipulation}
\begin{figure}[H]
    \centering
    \includegraphics[width=\textwidth]{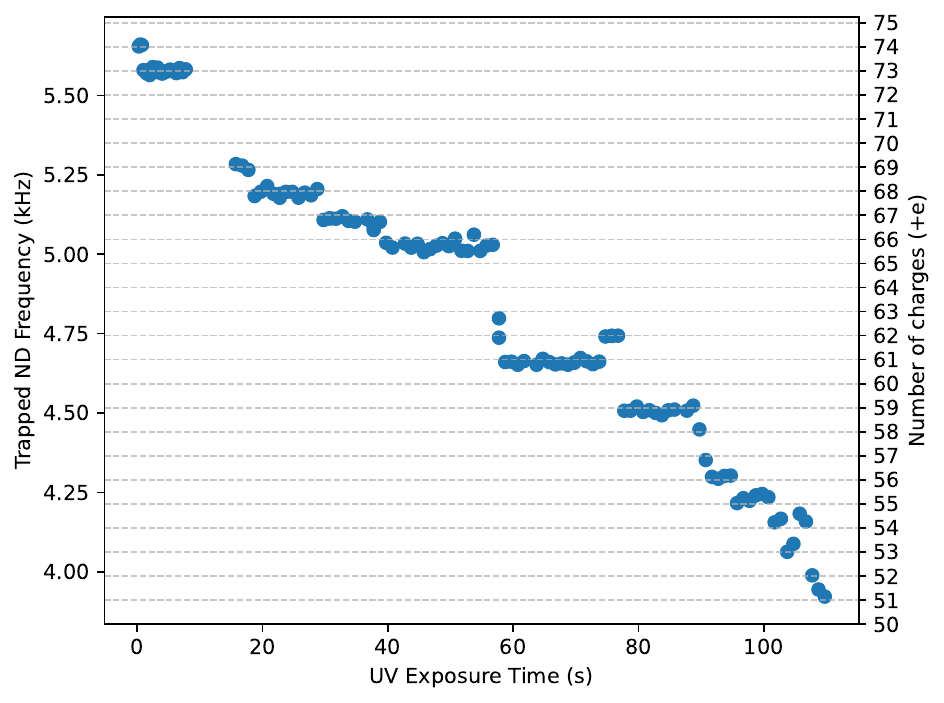}
    \caption{Single-electron charge manipulation using UV illumination. We plot the frequency vs. UV exposure duration for which exhibits discrete frequency steps. The particle was loaded via electrospray with positive voltage and illuminated with a UV LED at $263\pm 6\,$nm wavelength, focused using two lenses. The left $y$-axis shows the measured secular oscillation frequency of a levitated ND (250\,nm diameter) in a needle Paul trap at 0.5\,Torr pressure. The right $y$-axis displays the estimated net number of elementary charges, obtained from a fit to a discrete function of the form $f=\delta f\cdot N_e$, where $N_e$ is the number of electrons, and $\delta f$ is the discrete frequency step when adding/removing one electron. The optimal fit yields discrete frequency steps with multiples of $\delta f = 76.4$\,Hz, indicating single-electron charge steps. The UV illumination causes photoemission from the metallic trap electrodes, where the emitted electrons are subsequently captured by the positively charged ND particle. The stepwise frequency changes demonstrate precise control of individual electron charges on the levitated nanodiamond.}
    \label{fig:freq_vs_UVtime}
\end{figure}

We demonstrate the single charge manipulation of levitated charged NDs in a needle Paul trap using UV photoemission at a pressure of 0.5\,Torr. The experimental setup, which is described in detail in~\cite{skakunenko_strong_nodate} consists of a needle Paul trap into which particles are loaded using electrospray, with a positive voltage, such that the particles are assumed to be positively charged. The secular oscillation frequency of the particle is detected using split detection, and analyzed on a spectrum analyzer. We illuminate the trapping region using a UV LED with a measured wavelength of $263\pm 6\,$nm, focused by two lenses, and measure the frequency vs. UV exposure duration, which shows discrete frequency steps, indicating discrete charge change. The results are shown in Fig.\,\ref{fig:freq_vs_UVtime}. In this configuration, the UV illumination primarily causes photoemission from the metallic trap electrodes, such that some of the emitted electrons are subsequently absorbed by the positively charged ND particles.

The frequency is measured at a pressure of 0.5\,Torr, and the particle size is estimated to be 250\,nm in diameter. The measured frequency vs. UV exposure time is shown in Fig.\,\ref{fig:freq_vs_UVtime}. The right $y$-axis shows the estimated net number of elementary charges, obtained from a fit to a discrete function. The optimal fit yields jumps with multiples of $\delta f = 76.4$\,Hz, indicating discrete, single-electron charge steps in the particle.

\subsection{Fast neutralization of NDs using a UV laser}
\begin{figure}[H]
    \centering
    \includegraphics[width=\textwidth]{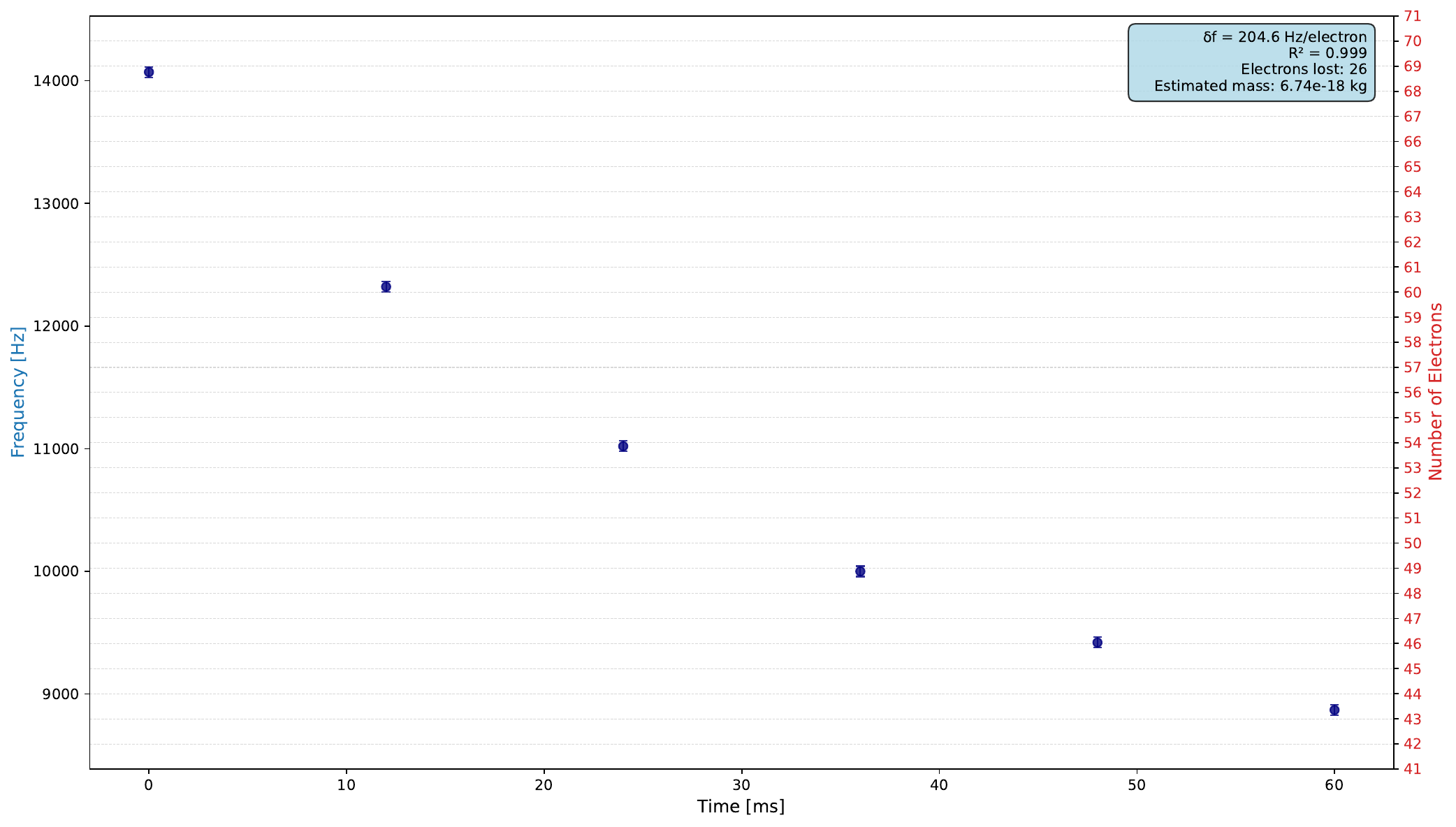}
    \caption{Fast neutralization of NDs using a UV laser. We plot the frequency vs. UV exposure duration. The particle was neutralized by using a 266\,nm UV laser, where the laser exposure duration was controlled using a mechanical shutter with an exposure duration of $\sim$12\,ms. The left $y$-axis shows the measured secular oscillation frequency of a levitated ND in a needle Paul trap at 0.2\,Torr pressure. The right $y$-axis displays the estimated net number of elementary charges, obtained from a fit to a discrete function of the form $f=\delta f\cdot N_e$, where $N_e$ is the number of electrons, and $\delta f$ is the discrete frequency step when adding/removing one electron. The optimal fit yields discrete frequency steps with multiples of $\delta f = 204.6$\,Hz, and an intial total charge of 69 electrons, which is consistent with the size of the particle. In the first pulse of duration 12\,ms we observe a frequency shift of $\sim$ 1750\,kHz corresponding to a net charge of 8.5 electrons, so that the rate of neutralization is $\sim$ 1.4\,ms per electron, which is approximately 100 times faster than the current state of the art~\cite{monteiro_force_2020} standing at 100\,ms per electron. As the neutralization is fast, we do not observe the discrete frequency steps of a single electron. This puts some uncertainty on the initial charge of the particle (or on the frequency shift due to one electron leaving the ND). Nonetheless, we can set a lower bound on the neutralization rate, just by assuming that the smallest step in the figure is one single electron change. This results in an initial charge of only 23 electrons, and the first step is now of $\sim$3 electrons, such that the initial neutralization rate is 4\,ms, which is still approximately 25 times faster than the current state of the art~\cite{monteiro_force_2020}.}
    \label{fig:fast_ neutralization}
\end{figure}

Here we report our results of fast neutralization of levitated charged NDs using a focused UV laser. The experimental setup is the same is described in Sec.\,\ref{sec:single_charge_manipulation}, with the addition of a focused 266\,nm laser beam for rapid neutralization. We measure the neutralization dynamics by monitoring the particle's oscillation frequency in the Paul trap as a function of laser exposure time. The laser unit is from Teem Photonics model SNU-02P-000, which provides an avarage output power of 7.7\,mW. This is a passively Q-switched Nd:YAG laser with a repetition rate of 9.2\,kHz, and a pulse duration of 0.5\,ns, which exploits the fourth harmonic generation of the fundamental wavelength of 1064\,nm and converts it to a wavelength of 266\,nm. The laser beam is focused using a lens with a focal length of 150\,mm, resulting in a spot size which was measured to be smaller than 200\,\(\mu\)m. The laser beam is aligned such that it is centerd between the two needles of the Paul trap, and does not illuminate the needles. We use a mechanical light shutter to control the laser exposure time, and meausre the oscailation frequency of the partilce vs. laser exposure duration. The results are presneted in Fig.\,\ref{fig:fast_ neutralization} and exhibit fast netrailzation.

\section{Discussion}
The results presented in this work address a significant research gap in the field of charged particle neutralization, particularly for levitated NDs. While previous studies have demonstrated neutralization techniques for silica particles, the application of UV photoemission to NDs has not been thoroughly explored. We provide clear evidence of neutralization of NDs using UV illumination in a Paul trap by measuring either the lifetime or frequency of the trapped NDs. We identify a wavelength for efficient photoemission at approximately 270\,nm (4.6\,eV). This value is within the range of previously reported work function measurements on diamond surfaces \cite{tafel_femtosecond_2019, Zhang_Abs_2023, mackie_work_1996, velardi_uv_2017}.
We measure the neutralization rate vs. particle size and observe a scaling law with $1/d^{1.3}$, which may shed light on the different photoemission mechanisms participating in the process. Finally, we achieve fast neutralization, faster by at least an order of magnitude than the state of the art. The results reported here confirm the feasibility of UV-based neutralization for levitated NDs and provide insight into the requirements for such neutralization systems, suggesting that a $266\,\mathrm{nm}$ laser with $10\,\mathrm{mW}$ power should be sufficient for fast charge manipulation and neutralization of a $75\,\mathrm{nm}$ levitated ND. Specifically, we verify that using a 266\,nm one can achieve neutralization at a rate of nearly one electron per ms. This work lays the foundation for further studies aimed at achieving rapid and precise charge control and neutralization under ultra-high vacuum conditions, a critical step toward realizing quantum interferometry experiments with NDs.
\section{Outlook}
Accurate charge control is essential for the implementation of matter-wave interferometry experiments with levitated NDs. In this work, we laid the foundation for the use of UV photoemission for fast neutralization of levitated charged NDs. Our next goal is to neutralize the NDs in an optical dipole trap, or a diamagnetic trap, where we can confirm zero net charge by applying a weak electric field and verifying no interaction, as planned for the ND SGI sequence, and implementing a feedback system to modulate the laser power. Another practical aspect for future work is testing charge control of positively charged NDs. Finally, several fundamental questions present themselves for investigation, such as the source of the emitted electrons during the photoemission process, the transport of charge on the ND surface, and the effect of the surface termination on the photoemission process.

\subsection{Single nanosecond laser pulse photoemission from NDs using a UV laser}
\begin{figure}[H]
    \centering
    \includegraphics[width=0.5\textwidth]{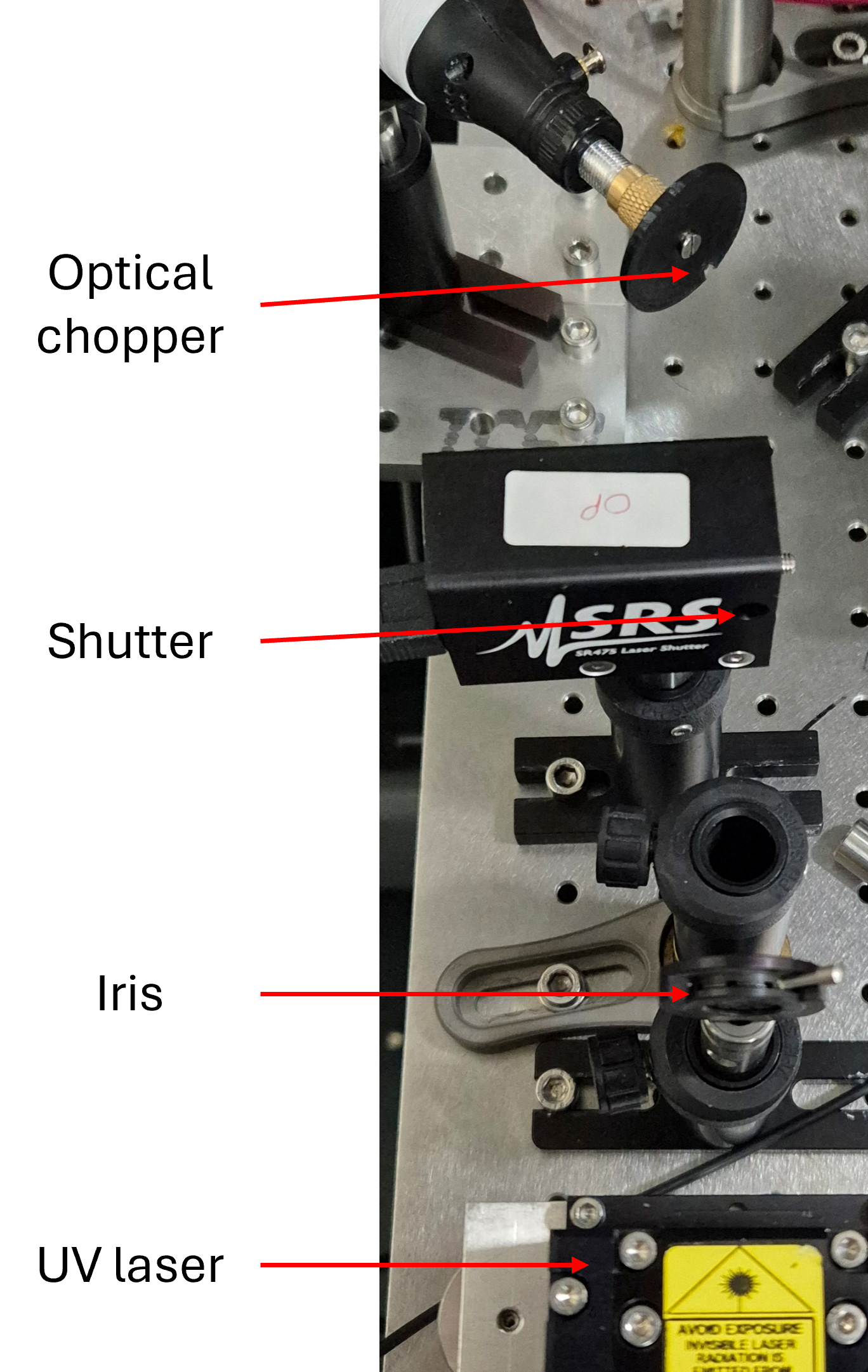}
    \caption{Setup of the single pulse picker. The beam is produced by a 266\,nm UV laser, with a repetition rate of 9.2\,kHz. An iris is used to improve the shape of the beam. The beam passes through a mechanical shutter (SRS SR475) with an exposure duration of 4\,ms, and through a homemade optical chopper, with a frequency of 250\,Hz and duty cycle of 1.3\%. This combination allows us to isolate a single laser pulse, resulting in an average of 0.7 laser pulses per shutter exposure.}
    \label{fig:optical_beam_picker}
\end{figure}

\begin{figure}[H]
    \centering
    \includegraphics[width=\textwidth]{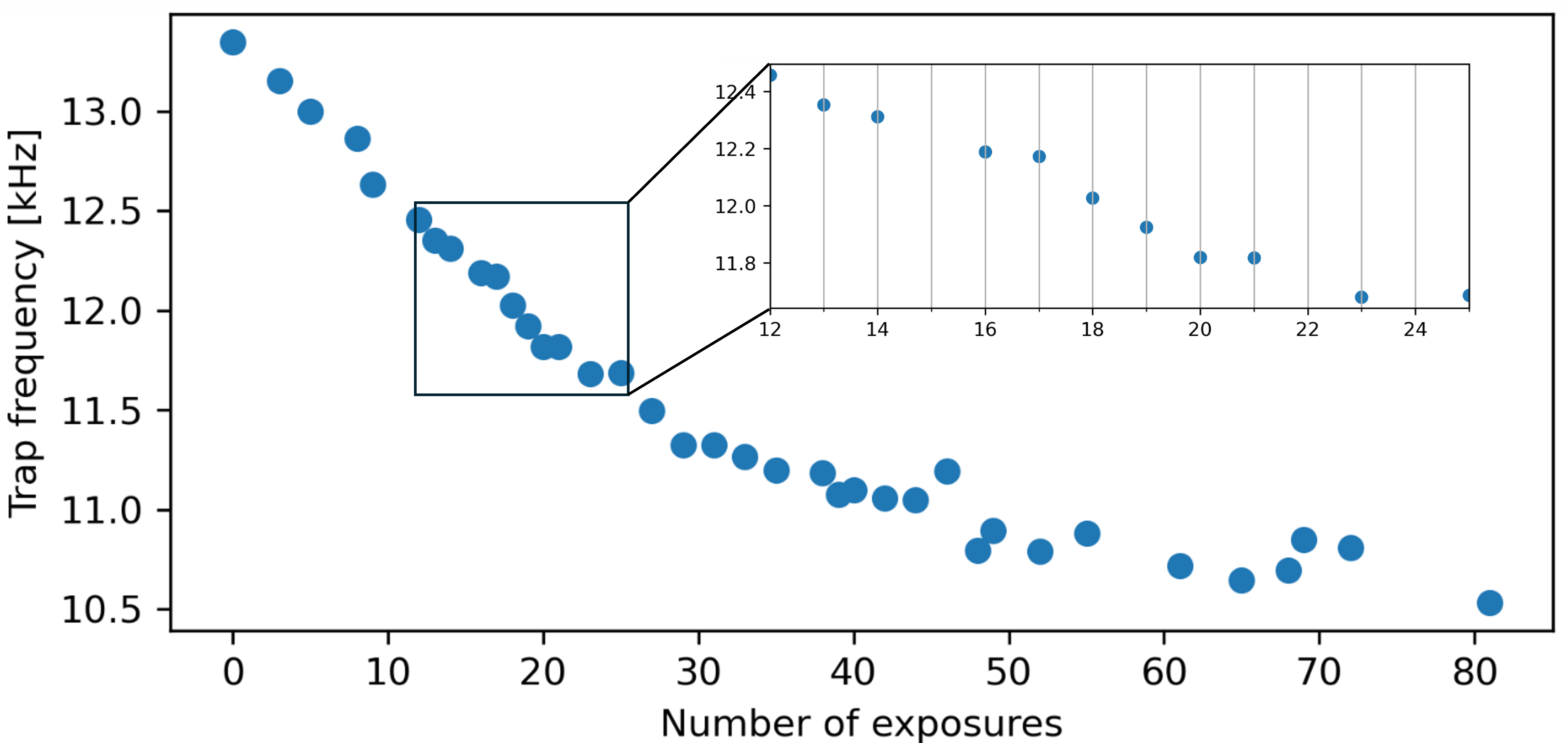}
    \caption{Photoemission of electrons from an ND due to a single nanosecond laser pulse. We isolate a single laser pulse using a combination of a mechanical shutter and an optical beam chopper. We plot the secular oscillation frequency vs. the number of shutter exposures. The data shows that one shutter exposure can result in a frequency step, indicating that a single nanosecond laser pulse can induce photoemission from the ND. The inset shows the range in which the phenomenon is prominent and easily visible.}
    \label{fig:single_pulse_photoemission}
\end{figure}

Here, we report preliminary results in which a a single nanosecond laser pulse induces photoemission of electrons from a levitated ND. To isolate one single laser pulse, we use a combination of a mechanical shutter and an optical beam chopper. The shutter is an SRS model SR475, which we measured to have an exposure duration of 2-4\,ms depending on the location of the beam relative to the shutter's knife, and we align the beam such that the exposure duration is $\sim$4\,ms. The optical beam chopper is homemade from a fast rotating motor (a battery powered rotary tool, rated at 16,000\,RPM (266.6\,Hz) and a measured rotation frequency of $\sim$250\,Hz) on which we mount a 3D printed disk with a diameter of 30\,mm and a single slit of $\sim$2.5\,mm width, which results in a duty cycle of $\sim$1.3\%. The optical chopper thus has an exposure duration of $\sim 50\,\mathrm{\mu s}$ every 4\,ms. This combination allows us to isolate a single laser pulse. We verify this by measuring on a fast photodiode an average of 0.7 laser pulses per shutter exposure, effectively creating a "single pulse picker". The apparatus of the shutter and optical chopper is shown in Fig.\,\ref{fig:optical_beam_picker}. Using this single pulse picker, we measure the frequency of the particle in the Paul trap as a function of the number of shutter exposures and observe the frequency shifts. The results are shown in Fig.\,\ref{fig:single_pulse_photoemission}, where some of the single exposures result in a distinct frequency shift, indicating that a single nanosecond pulse can induce photoemission of an electron from the ND.

\section*{Acknowledgments}
We thank the BGU Atom-Chip Group support team, especially Menachem Givon, Zina Binstock, Dmitrii Kapusta, and Yaniv Bar-Haim for their support in building and maintaining the experiment. Funding: This work was funded by the Gordon and Betty Moore Foundation (\href{https://doi.org/10.37807/GBMF11936}{doi.org/10.37807/GBMF11936}), and the Simons Foundation (\href{https://www.simonsfoundation.org}{MP-TMPS-00005477}).



\end{document}